# Field emission characteristics of carbon nanotubes according to the high energy ion irradiation

P.M. Koinkar[1,*], A. Kumar[2], D.K. Avasthi[3], M.A. More[4], Ri-ichi Murakami[5]

[1,*]*Center for International Cooperation in Engineering Education (CICEE), University of Tokushima, Tokushima-770 8506, Japan*

[2]*Institute for Experimental and Applied Physics, University of Regensburg Regensburg-93040, Germany.*

[3]*Materials Science Group, Inter University Accelerator Centre, PB 10502, New Delhi-110067, India.*

[4]*Department of Physics, University of Pune, Pune-411007, India*

[5]*Department of Mechanical Engineering, University of Tokushima, Tokushima 770 8506, Japan*

-----------------------------------------------------------------------

**\* Author to whom correspondence should be addressed**

koinkar@tokushima-u.ac.jp

**Abstract:**

The field emission properties of ion irradiated multiwalled carbon nanotubes (MWCNTs) and doublewalled carbon nanotubes (DWCNTs) have been studied. The carbon nanotubes synthesized by a chemical vapor deposition method were irradiated by high energy (90 MeV) Au ions with different ion fluence from $4 \times 10^{11}$ to $1 \times 10^{13}$ ions/cm$^2$. After ion irradiation, the field emission properties of MWCNTs and DWCNTs were greatly influenced. The change in the emission characteristics is due to structural defects caused by the high energy ion irradiation. The emission characteristic of MWCNTs was improved and turn-on field decreased from 5.43 to 3.10 V/μm by ion irradiation. Noticeable improvement in emission characteristics of MWCNTs was observed at a fluence of $1 \times 10^{13}$ ions/cm$^2$. The emission characteristics of DWCNTs deteriorated and the turn-on field was increased from 2.44 to 7.76 V/μm. This results show the distinctly different behavior of ion irradiated MWCNTs and DWCNTs.

**Introduction:**

The carbon nanotubes (CNTs) have attracted much attention because of their high aspect ratio, excellent electrical conductivity, chemical stability and mechanical robustness [1-4]. Among the various applications due to their unique and excellent properties, field emission displays (FEDs) and other vacuum microelectronics devices are being considered for the realistic applications area because of their high aspect ratio leading to high electric field enhancement and low operating voltage [5-7].

There have been great advances in fundamental properties and applications of the field emission from CNTs. Especially, enormous efforts in controlling and modifying their unique structure and remarkable properties induced many kinds of field emission applications. It is well known that surface modification of CNTs is preferable to obtain the high performance field emitters. According to previous works, the improvement in the emission characteristics of CNT could be achieved by various surface treatments such as plasma treatment, laser irradiation, focused ion beam and ion irradiation [8-12].

The effect of ion irradiation on the solid surface has been extensively studied due to its scientific and technological importance. Ion irradiation can change the physical and chemical properties as well as modify the morphology of solid surface because of lattice damage, ion sputtering, ion induced diffusion and chemical reaction which either change $sp^2$ content of G-phase or decrease $sp^3$ content of D-phase. The ion irradiation creates local amorphous region and recrystallises the lattice. The possible surface modification depends on the type of ion irradiated material.

Despite the technical importance and it's unique features, there have been very few reports about ion irradiation effects on CNTs, especially surface changes and structure destruction. Several studies have been carried out to change the morphology and the properties of CNTs using ion irradiation. Wei et al. performed the 50 keV $Ga^+$ ion irradiation on MWCNTs and showed the formation of highly ordered pillbox like nanocompartment [13]. The nanotubes defects could be formed by introducing the irradiation with highly charged particles. It has been shown that both electron and heavy ion irradiation could modify the structure and the dimension of the CNTs [14,15]. Zhu et. al have shown that energetic $Ar^+$ ion irradiation generates dangling bonds (vacancies) on the surface of CNTs [16]. Recently some research groups reported that ion irradiation is an effective tool to modify the surface and dimension of CNTs [17,18].

Interestingly, it has been reported that low energy ion irradiation can improve the field emission characteristics of CNTs. Kim et. al successfully demonstrated that the CNTs with Ar ion irradiation exhibited better field emission characteristics due to the straightening of CNTs and the increase in the number of defects [11,19]. However, in most cases, low Z ion mainly has been used with low energy of about 30 keV to few MeV. We considered that, in case of ion irradiation, the structural modification is generated as a function of energy and mass of incident ion. Thus, we use high Z ion with high energy to get more concrete effects of ion irradiation. In this work, we investigated the effect of high energy ion irradiation on the field emission characteristics of MWCNTs and DWCNTs. Our purpose is to modify the morphology and structure of CNTs and to evaluate the emission characteristics of CNTs with ion irradiation treatment.

**Experimental:**

## 1) Synthesis of MWCNT :

The synthesis of as-grown MWCNTs was carried out using chemical vapor deposition (CVD). The n-Si substrate, on which Fe catalyst with thickness of 10 nm was sputtered, is used to grow the MWCNTs. The acetylene ($C_2H_2$) and ammonia ($NH_3$) were used as source gases with a flow rate of 30 sccm and 300 sccm for 10 minutes at 750 $^0C$, respectively. The $NH_3$ pretreatment was employed prior to the growth of MWCNTs.

## 2) Synthesis of DWCNT :

As-grown DWCNTs were synthesized by CVD using Fe-Mo / MgO catalysts. The synthesis was carried out in a quartz tube reactor (70 mm i.d., and 700 mm long) mounted in a tube furnace. Fe-Mo catalyst (~200 mg) was put into a quartz boat at the center of the reactor tube and then quartz tube was heated up to 900 $^oC$ in an Ar atmosphere. The DWCNTs were grown with gas flow of $CH_4$ (300 sccm) and a mixture of Ar (500 sccm) and $H_2$ (100 sccm) for 20 minute at 900 $^oC$.

## 3) Ion irradiation

The ion irradiation was carried out using 90MeV Au ion at room temperature from 15 UD pelletron accelerator. The ion current was about 1 pnA (particle nanoampere) which is equivalent to ~ 6.25x $10^9$ ions/s. The samples were mounted inside the irradiation chamber which was evacuated at ~ 1x $10^{-6}$ mbar. The samples were irradiated with different ion fluence of 4x $10^{11}$, 4x $10^{12}$, 1x $10^{13}$ ions/cm$^2$.

## 4) Field emission measurement of MWCNTs and DWCNTs

For the field emission measurement from MWCNTs and DWCNTs, we have fabricated the field emitters as follows. In case of MWCNTs, as-grown MWCNTs on n-Si/Fe(10nm) is used as cathode. For DWCNTs, field emitters were fabricated by depositing DWCNTs on Ag paste coated n-Si substrate using a sieving method with brush. The substrate was dried in air for 1 hr, followed by baking at 80 ºC for 1 hr for keeping good adhesion and ohmic contact property between DWCNTs and substrates.

The field emission studies of as-grown and $Au^+$ ion irradiated CNTs were carried out in a vacuum chamber with a base pressure ~ 2 x $10^{-7}$ torr. The CNT films were used as cathode and stainless steel plate (Φ = 5 mm) was used as anode. The distance between anode and cathode was 300 µm and the measured emission area was 0.19625 $cm^2$. Emission current was measured with a Keithley 6517 A and DC power was supplied with a constant voltage and current controller (HCN140-3500 of Hochspannungs-Netzgerät: 0 – 3.5 kV, 0 - 40 mA). The macroscopic field was defined as the values of applied voltages divided by the gap between the anode and cathode.

**Results and discussion:**

The MWCNTs, used in present work, were prepared by CVD method and are about 30 nm diameter and 4.5 µm in length. The typical results of emission current density (J) and electric field (E) for as-grown and ion irradiated MWCNTs are presented in Fig. 1 (a). In this work, the MWCNTs were irradiated with by $Au^+$ ion with 90 MeV at different ion fluence of $4x10^{11}$, $4x10^{12}$, $1x10^{13}$ ions/$cm^2$. The J-E characteristic shows very different results in ion irradiated MWCNTs and DWCNTs. Here the turn-on field and threshold field are defined as the electric field at which emission current reaches to 3 µA/$cm^2$ and 1 mA/$cm^2$ respectively. It can be seen that the turn-on field and threshold field of as-grown MWCNTs are found to be 3.52 and 5.41 V/µm, respectively. The emission of electrons comes from the CNT tips only. The field emission properties are relatively steady after voltage cycling.

After the irradiation, at ion fluence of 4x10$^{11}$ ions/cm$^2$, the turn-on field and threshold field are measured to be 3.66 and 5.31 V/µm, respectively. These values are almost comparable with as-grown MWCNTs sample. But, the careful observation J-E characteristics of MWCNT at ion fluence 4x10$^{11}$ ions/cm$^2$ reveals that the emission current density increases slightly above the threshold field because of irradiation as compared to as-grown MWCNTs. In general, electrons transport along the NTs and could emit from the CNT tips. However, if the defects are created on the walls of NTs due to irradiation, then electron could emit from there as well. The energetic ions strike the CNTs tips and protruding CNT tips are destroyed gradually. Subsequently, the non-protruding CNTs tips start to emit because the number of sites was increased after the irradiation at ion dose of 4x 10$^{11}$ ions/cm$^2$. As a result of this, the improvement in the emission current above the threshold field was observed as compared to as-grown MWCNTs.

The turn-on field and threshold field are decreased further from 2.69 to 2.01 V/µm and 3.92 to 3.10 V/µm respectively for an increase in ion fluence of 4x10$^{12}$, 1x10$^{13}$ ions/cm$^2$. The noticeable improvement in the emission characteristics of MWCNTs was observed for ion fluence of 1x10$^{13}$ ions/cm$^2$. The highest emission current density of 1 mA/cm$^2$ at electric field as low as 3.10 V/µm has been seen for ion fluence of 1x10$^{13}$ ions/cm$^2$. Hence the ion irradiation of MWCNTs with different ion fluence exhibits the enhancement in the emission current density and reduction in the threshold field. The reduction in the electric field may be attributed to the surface modification in NTs. The incident ion creates the defects into the outer NTs wall of the surface. These defects on the outer wall could be responsible for increase in the emission sites favorable for field emission. Although the electrons emit from CNT tips before ion irradiation, majority of electrons come from the defects induced walls of CNTs after the ion irradiation. As the ion fluence increases, more protruding features could be generated by the impact of highly energetic ion. Each ion may produce a protrusion since incident ion pass through surface which causes damage protrusion. Thus the increase in the number of defects and surface modification due to ion irradiation plays a vital role in exhibiting better field emission characteristics.

According to the Fowler-Nordheim (F-N) equation, the field emission current density (J) is given by

$$J = A(\beta^2 V^2/\phi d^2) \exp(-B\phi^{3/2} d/\beta V)$$

where $A$ (= 1.54 x $10^{-6}$ (AV$^{-2}$ eV) and $B$ (= 6.83 x $10^9$ (VeV$^{-3/2}$ Vm$^{-1}$) are the proportionality constants, $\beta$ is the field enhancement factor, $\phi$ is the work function, E =($V/d$), $d$ is the distance between anode and cathode, $V$ is applied voltage [20,21]. The $\beta$ can be calculated as $\beta = -6.83 \times 10^9 \times \Phi^{3/2}/m$, where $m$ is the slope obtained from the F-N plot. Figure 1(b) shows F–N plots of as-grown and ion irradiated MWCNTs. The field enhancement factor $\beta$ is found to be increased from 833 to 2071. The increase in the field enhancement factor can be explained as follows. After ion irradiation, most of the CNTs have sharp tips and these sharpened tips may act as sharp field emitters. Moreover, the CNTs may be well separated on the surface which reduces the screening effect between two neighboring tips.

The MWCNTs films irradiated with different ion fluence have lower threshold field as compared to as-grown film. This is attributed to the defects generation on the wall of CNTs. The better emission may be due to existence of the defects and these defects also contributes to the electron field emission as the tips of CNTs. A detailed explanation is given below. It is believed that the ion irradiation effects in CNTs are different from those occur in common crystals. During the ion irradiation treatment, ion preferentially attacks the protruding CNTs tips and wall of the CNTs and generates planer defects on the walls of CNTs. When an energetic ion strikes on the CNTs, it transfers energy to one–to-three atoms in the uppermost shell by creating many primary carbon recoils and a single-multiple atoms vacancy. It is shown that, for MWCNTs, ion-irradiation produced vacancies in graphene networks, causes removal of many carbon atoms from $sp^2$ bonded carbon nets of NTs and intershell links due to the saturation of dangling bonds at carbon interstitials and atoms nearby the vacancies [ 13, 22-26]. The number of defects CNTs can be created if the incident ion energetic enough to displace the carbon atom. Upon ion irradiation, the number of structural defects and amorphous carbon and increase the degree of disorder produced due to the mechanical impact high energetic ion. Pomoell et al. performed the simulation about high energy ion irradiation on MWCNTs [27]. They explained that the damage created in the NTs quickly increases with irradiation fluence. But, at higher ion fluence, the number of defects saturates, because at certain irradiation fluence, the system has reaches the degree of disorder so large that incoming ion will not result in any further significant change on the wall of CNTs. Recently, there are few reports on the ion irradiation of MWCNTs which explains that the irradiation induced defects appear hillock-like protrusions on the CNTs wall [24,28-31] and number of defects with formation of nanocompartment with bamboo-like structure inside the tube after ion irradiation [32].

However, the effect of plasma ion irradiation for the improvement of field emission properties has been studied by several researchers [8,9,33-42]. It is proved that the improvement in the field emission characteristics occurs for MWCNTs irradiated with $Ga^+$ ion with energy 30 keV at low ion fluence. The enhancement in field emission characteristics is due to ion irradiation, which creates more the emission site and , presumably, lowers the work function, by dangling bonds [9]. Moreover, the MWCNT were irradiated by $Ar^+$ plasma treatment give rise to better field emission properties. After Ar ion plasma treatment, Ar ion penetrates the layer of CNTs which leads to formation of dangling bonds. The dangling bonds on the top of the surface works as emission site to enhance the field emission [8]. The oxygen plasma treatment has been employed to improve the field emission characteristics of MWCNT. The density of CNTs decreases as results of CNTs destruction after plasma treatment. The graphitic structure of the outer wall of the NTs was destroyed and the tips of the CNTs were sharpened by plasma treatment. The structural defects were serve as an emission sites. Thus the defects and sharpened tip are the main reasons for enhanced field emission properties after $O_2$ plasma treatment [33].

According to some researchers group, the enhancement in the field emission characteristics after plasma treatment is due the change in surface morphology, formation of nanoparticle and structural defects [34-36]. Similar studies have been carried to improve the field emission properties of CNTs with Ar, $O_2$, $H_2$ plasma treatments . After the plasma treatment, the amorphorism of CNTs structure was produced with few defects such as dangling bonds, interlayer cross-linking, $sp^3$ defects and the structural changes in surface due to high energetic ion bombardment from the plasma. Thus the improvement in field emission performance of CNTs was attributed to the surface reconstruction, many defects, and sharpening of emitter [37,38]. Recently similar reports appeared on the enhancement in the field emission properties of aligned MWCNTs after laser irradiation and plasma ion (like $CF_4$ and $O_2$ and Ar) irradiation [39-42]. According to these reports, the possible factors for increase in field emission properties of CNTs after ion bombardment by plasma treatment are structure modification, reduction in surface density, generation of large defects, formation of sharp CNT tips, reduction of workfunction of CNTs film. The above all results are based on the plasma and laser ion irradiation with low energy about few keV. However, present study shows the field emission improvement after high energy (90 MeV) ion irradiation. We suggest that the enhancement in the current density and decrease in the threshold field is probably comes from the generation number of defects into the outer NTs wall of the surface. The presence of the defects on the surface of CNTs greatly influences the structural transformation in the CNT structure under high energy ion irradiation. The increase in the structural defects is the origin of generating more emitting site which easily emits the electrons and also contributes for better field emission characteristics of MWCNTs films. Thus the change in electronic structure ( like defects on the wall of the NTs) caused by ion irradiation is effective way to enhance the field emission properties of CNTs.

The emission current density (J) and electric field (E) characteristics for as-grown and ion irradiated DWCNTs are presented in Fig. 2 (a). In this work, the DWCNTs were irradiated by 90 MeV Au ion different ion fluence of $4x10^{11}$, $4x10^{12}$, and $1x10^{13}$ ions/cm$^2$. The J-E characteristic shows very different results as compared to irradiated MWCNTs with increase in ion fluence. The turn-on field and threshold field of as-grown

DWCNTs are found to be 1.46 and 2.44 V/µm respectively. After irradiation at ion dose of $4\times10^{11}$ ions/cm$^2$, the turn-on field and threshold field are increased to be 1.63 and 2.85 V/µm, respectively. Also, the current density has been decreased with increase in ion irradiation. These results are in contrary to those reported above for MWCNT. In case of MWCNTs, the change in ion fluence leads to the significant improvement in emission current density and the enhancement in field emission characteristics and this is because of defects generated by irradiation. But, in case of DWCNTs, we observed the degradation of emission current which may be attributed to severe structural damages created during ion irradiation. At ion fluence of $4\times10^{12}$ ions/cm$^2$ and $1\times10^{13}$ ions/cm$^2$, the turn-on fields are increased from 3.70 to 4.44 V/µm, while threshold fields are also increased from 6.35 to 7.76 V/µm. Figure 2(b) shows F–N plots of as-grown and ion irradiated MWCNTs. The field enhancement factor $\beta$ found to be decreased from 1386 to 920. This clearly shows the reduction in emission current density of DWCNTs with ion irradiation treatment. Hence the ion irradiation of DWCNTs with different ion fluence exhibits the degradation in the emission current density and increase in the threshold field which deteriorate the emission characteristics.

The reason for degradation in the emission characteristics and increase in the electric field of DWCNTs is quite different from those of MWCNTs. After ion irradiation with high ion fluence, the DWCNTs quickly undergoes structural modification as compared to MWCNTs. During the high energetic ion bombardment, the DWCNTs are subjected to structural change because of continuous collisions of Au ions with the carbon atoms in DWCNTs. At the end of the ion irradiation, the structural rearrangement into graphene layers of DWCNTs took place. The Au ion pass through the uppermost graphitic layer of the DWCNTs and ion can penetrates the second graphitic layer as well. Hence, ion irradiation leads to deterioration and finally amorphorisation of the CNTs wall resulting in a noticeable loss mechanical strength to the bundles of DWCNTs. Thus, the mechanical damage results in the severe destruction of DWCNTs. Since DWCNTs exhibit less mechanical strength than MWCNTs, the mechanical damage caused by the high ion irradiation may not be resisted by DWCNTs which leads to amorphorization. Therefore, the graphite structure of the wall of DWCNTs may destroyed. The possible justification is that, since the DWCNTs has only two graphene layers, ion irradiation causes more damage on the walls of NTs. This will generates defects and original graphitic structure may be destroyed. The mechanical damage quickly increases with increase in ion fluence. As ion fluence increases, there will be more and more destruction of DWCNTs caused by mechanical impact of accelerated high energy ion irradiation. Hence more structural damage of DWCNTs results in degradation of field emission characteristics.

**Conclusion:**

The field emission characteristic of MWCNTs and DWCNTs are significantly influenced by high energy $Au^+$ ion irradiation. The present study shows the noticeable improvement in the emission current from MWCNTs and degradation in emission current from DWCNTs after high energy Au ion irradiation treatment. The enhancement in emission characteristics of MWCNTs is due to defects created on the wall of CNTs caused by $Au^+$ ion irradiation. On the other hand, in case of DWCNTs, the degradation in emission current density is due to more destruction of DWCNTs caused by higher excessive defect generated by high energy ion. This clearly suggests that

MWCNTs has more radiation resistance than DWCNTs. This study gives a contrasting behavior of DWCNTs and MWCNTs under high energy ion irradiation.

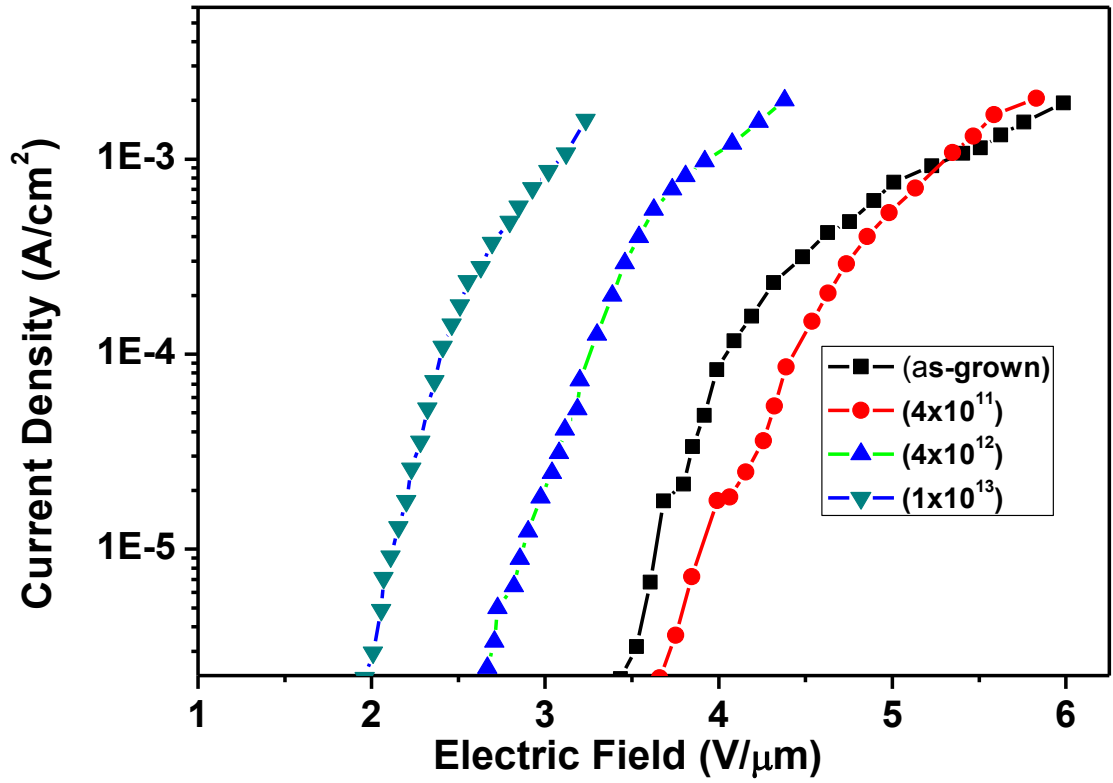

Figure 1 : (a) J-E characteristics of as-grown and irradiated MWCNTs

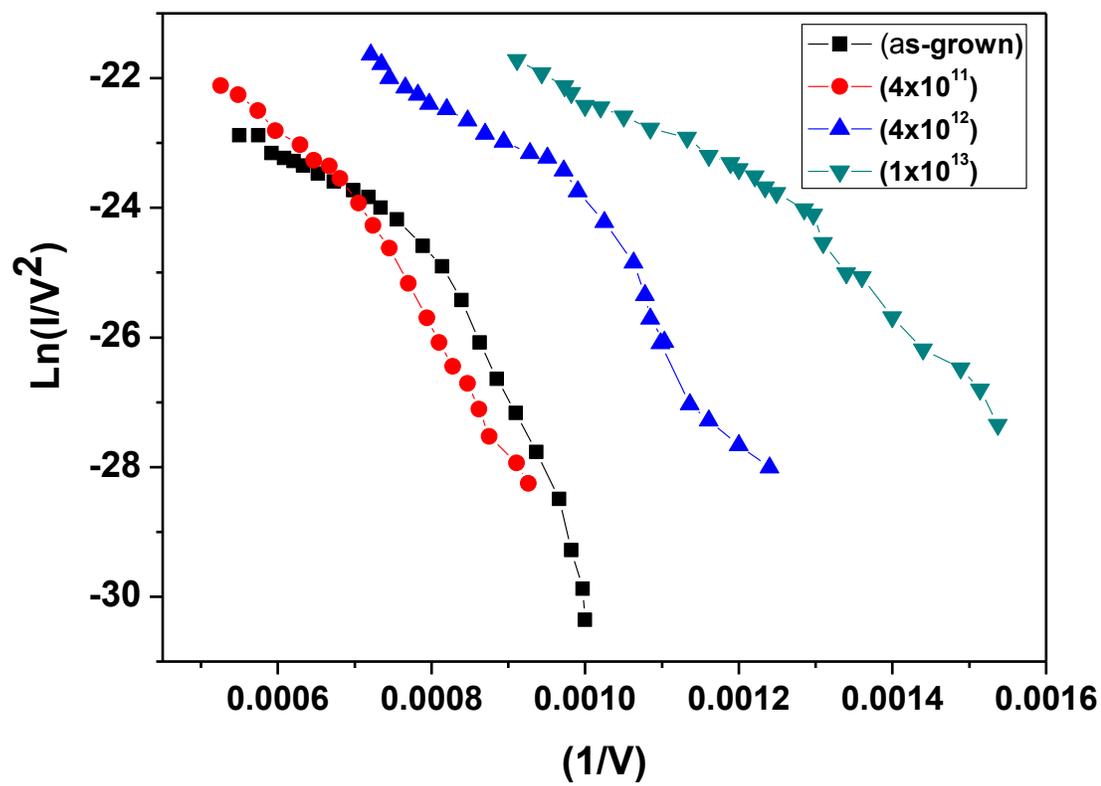

Figure 1 : (b) Corresponding F-N plots of as-grown and irradiated MWCNTs

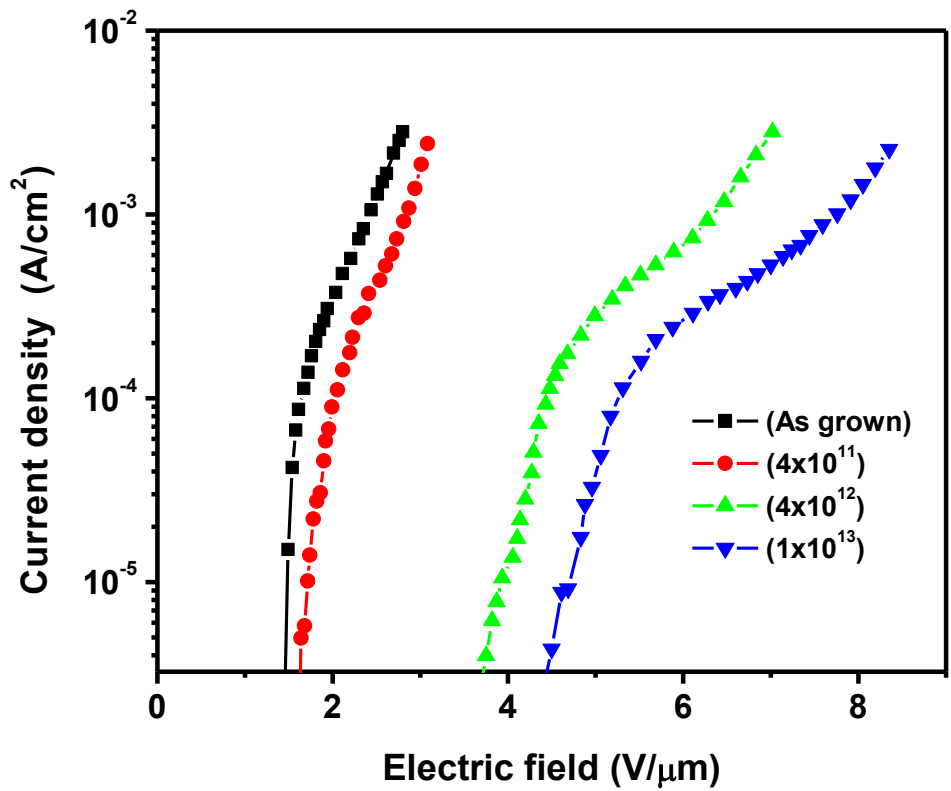

Figure 2 : (a) J-E characteristics of as-grown and irradiated DWCNTs

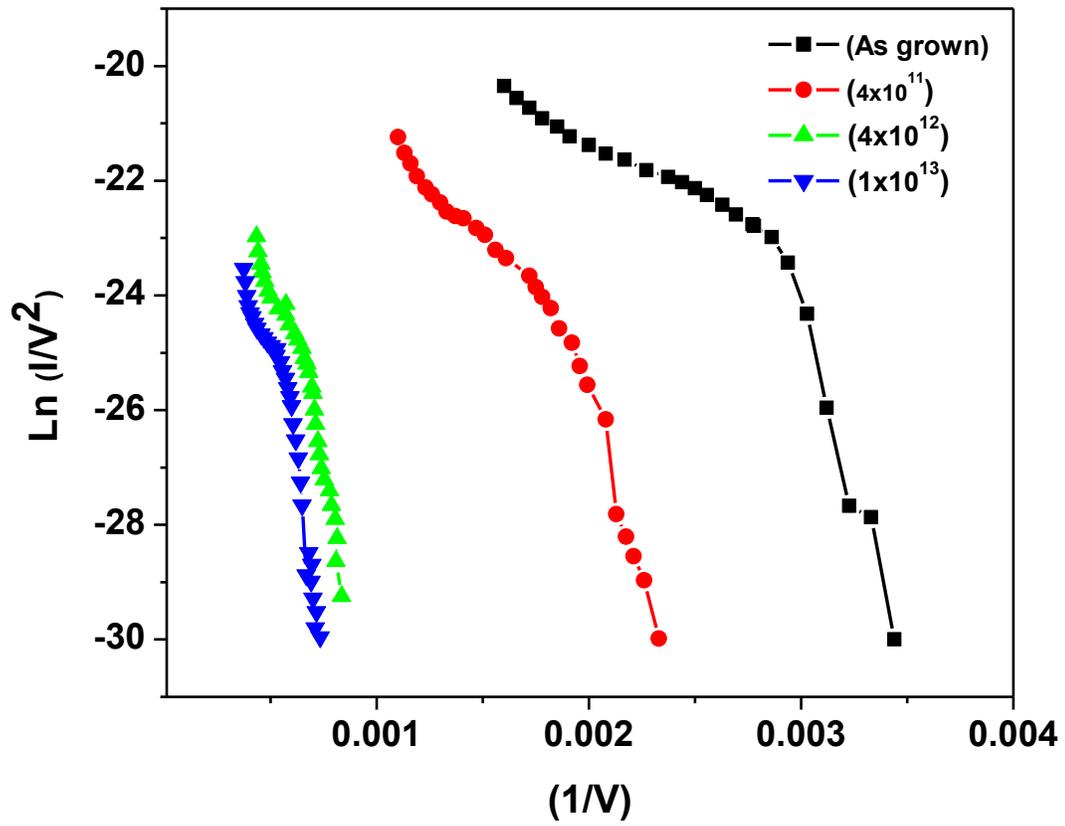

Figure 2 : (b) Corresponding F-N plots of as-grown and irradiated DWCNTs